\documentstyle[preprint,aps]{revtex}

\newcommand{\krc}{k r_c}
\newcommand{\krcp}{k r_c\pi}

\begin{document}

\setlength\baselineskip{20pt}

\preprint{\tighten\vbox{\hbox{CALT-68-2250}\hbox{hep-ph/9911457}}}

\title{Phenomenology of a Stabilized Modulus}

\author{Walter D. Goldberger\footnote{walter@theory.caltech.edu} and Mark B. Wise\footnote{wise@theory.caltech.edu}}
\address{\tighten California Institute of Technology, Pasadena, CA 91125}

\maketitle

{\tighten
\begin{abstract}
We explore the phenomenology of a stabilized modulus field in the Randall-Sundrum scenario.  It is found that if the large separation between branes arises from a small bulk scalar mass then the modulus (i.e. radion) is likely to be lighter than the lowest Kaluza-Klein excitations of bulk fields, and consequently may be the first direct signature of the model.  Four-dimensional general covariance completely determines the couplings of the modulus to Standard Model fields.  The strength of these couplings is determined by a single parameter which is set by the TeV rather than the Planck scale.
\end{abstract}}
\vspace{0.7in}
\narrowtext

\newpage

Although the Standard Model of strong and electroweak interactions successfully accounts for most experimental observations to date, it has several unattractive features that suggest the presence of new physics at high energies.  One of these features is the gauge hierarchy problem, which refers to the vast disparity between the weak scale and the Planck scale.  In the minimal Standard Model, this hierarchy is unnatural since it requires a fine tuning order by order in perturbation theory.  Several extensions have been proposed to avoid this fine tuning, such as low energy supersymmetry~\cite{SUSY} and technicolor~\cite{tech}.

More recently it has been proposed that the introduction of compactified extra spatial dimensions may also provide a solution to the hierarchy problem\cite{xdim}.  In their simplest form, these scenarios assume that spacetime is the product of a four-dimensional Minkowski space and a compact $n$-manifold.  While gravity can propagate freely through the extra dimensions, Standard model fields are confined to the four-dimensional spacetime.  Observers in this three-dimensional wall, or ``3-brane,'' measure an effective Planck scale $M_{Pl}^2 = M^{n+2} V_n,$ where $M$ is the fundamental Planck scale and $V_n$ is the volume of the extra dimensions.  If $V_n$ is large enough, $M$ can be of order the weak scale.  The hierarchy problem is then reformulated as the dynamical question of finding a mechanism that stabilizes the size of the extra dimensions.

Randall and Sundrum considered a different scenario that does not require large extra dimensions~\cite{RS1}.  Their model consists of a single $S^1/Z_2$ orbifold extra dimension with 3-branes residing at the boundaries of the spacetime.  The brane tensions together with a bulk cosmological constant give rise to a non-factorizable geometry with metric 
\begin{equation}
\label{eq:metric}
ds^2 = e^{-2 \krc |\phi|}\eta_{\mu\nu} dx^\mu dx^\nu - r_c^2 d\phi^2,
\end{equation}
where $k$ is a parameter of order $M,$ $r_c$ parametrizes the radius of the fifth dimension, and $x^\mu$ are Lorentz coordinates on the four-dimensional surfaces of constant $\phi\in [-\pi,\pi].$   The points $(x,\phi)$ and $(x,-\phi)$ are identified, and the two 3-branes are located at the orbifold fixed points, $\phi=0$ and $\phi=\pi.$  It can be shown that the effective four-dimensional Planck scale is given by 
\begin{equation}
M_{Pl}^2=\frac{M^3}{k}[1-e^{-2kr_c\pi}],
\end{equation}
so that even for large $kr_c$, $M_{Pl}$ is order $M.$  Because of the exponential factor in the spacetime metric, a field confined to the brane at $\phi=\pi$ with mass parameter $m_0$ has a physical mass $m_0 e^{-\krcp}.$  If $kr_c$ is around 12, the weak scale is dynamically generated on this ``visible'' brane from the scale $M,$ which is of order the Planck mass.  The basic setup of~\cite{RS1} is analogous to the Horava-Witten scenario~\cite{HW,mtheory} which arises in $M$-theory.  Supergravity and string theory realizations of the Randall-Sundrum model were suggested in~\cite{supergrav,verlinde}.  Variants of the basic scenario of~\cite{RS1} can be found in~\cite{others}.  An alternative which does not require the presence of a negative tension brane appears in~\cite{LR}. 

The scenario of~\cite{RS1} has several distinctive phenomenological consequences.  For example, Kaluza-Klein gravitational modes have masses given by the TeV scale and couplings to visible brane matter that are suppressed by a TeV~\cite{RS1,RS2}.  The implications of this for collider experiments have been studied in~\cite{DHR1}.  A similar pattern occurs for the Kaluza-Klein modes of other bulk fields, such as scalars~\cite{GW1} and bulk gauge fields~\cite{gauge}:  even if a bulk field has a mass which is of order the Planck scale, its low-lying Kaluza-Klein excitations have masses set by the TeV scale. 

The model presented in~\cite{RS1} contains a four-dimensional massless scalar, known as the modulus or radion, which determines the parameter $r_c$ in Eq.~(\ref{eq:metric}).  This field appears as one of the $\phi$-independent fluctuations about the background geometry.  Including these fluctuations, Eq.~(\ref{eq:metric}) becomes
\begin{equation}
\label{eq:fluc}
ds^2 = e^{- 2 k |\phi| T(x)} g_{\mu\nu}(x) dx^\mu dx^\nu - T^2(x) d\phi^2,
\end{equation}
where $g_{\mu\nu}$ is the four-dimensional graviton and $T(x)$ is the modulus field.  The orbifold symmetry excludes the propagation of a four-dimensional massless vector fluctuation.  A Kaluza-Klein reduction of the five-dimensional Einstein-Hilbert action for this metric leads to the following effective action for the massless modes $T(x)$ and $g_{\mu\nu}(x):$
\begin{equation}
\label{eq:phi}
S= 2 M^3\int d^4x d\phi\sqrt{-g} e^{-2 k|\phi| T} [6 k |\phi| \partial_\mu T \partial^\mu T - 6 k^2 |\phi|^2 T \partial_\mu T \partial^\mu T + T R],
\end{equation}
where $R$ is the Ricci scalar constructed from the metric $g_{\mu\nu}$.  After the $\phi$ integration there is a cancellation between the first two terms in Eq.~(\ref{eq:phi}) and only the part that depends on the exponential of $T$ remains:
\begin{equation}
S= \frac{2 M^3}{k} \int d^4x \sqrt{-g} \left(1-e^{-2 k\pi T}\right)R + \frac{12 M^3}{k}\int d^4 x \sqrt{-g}\partial_\mu \left(e^{-k\pi T}\right) \partial^\mu \left(e^{-k\pi T}\right).
\end{equation}
As we will see, this result implies that contrary to the claim of Ref.~\cite{RS1}, if $\krc\sim 12$, the modulus is coupled to visible brane matter with TeV rather than gravitational strength.  Defining $\varphi = f \exp(-k\pi T)$ with $f=\sqrt{24 M^3/k},$ we arrive at
\begin{equation}
\label{eq:action}
S= \frac{2 M^3}{k} \int d^4 x \sqrt{-g}{}\left(1- (\varphi/f)^2\right) R +\frac{1}{2}\int d^4 x\sqrt{-g} {}\partial_\mu \varphi \partial^\mu \varphi,
\end{equation}
so the original Randall-Sundrum scenario contains a massless scalar whose couplings to matter are set by the TeV scale.  This is clearly at odds with observation.  Furthermore, Eq.~(\ref{eq:action}) contains no dynamics that could stabilize $\varphi$ and give $T$ its desired VEV.  The proposal of~\cite{RS1} cannot be considered a complete resolution of the hierarchy puzzle until some additional dynamics to stabilize the modulus is specified.  It was shown in~\cite{GW2} that the presence of a bulk scalar propagating on the background solution of Eq.~(\ref{eq:metric}) can generate a potential $V(\varphi)$ that stabilizes the modulus (this scenario was generalized in~\cite{DFGK} to include the effects of the scalar on the background geometry).  The minimum of $V(\varphi)$ can be arranged to yield the desired value of $kr_c$ without extreme fine tuning of parameters.\footnote{However, to ensure a flat geometry on the branes, it is  still necessary to set the four-dimensional cosmological constant to zero.  Relaxing this constraint leads to bent brane solutions of the five-dimensional Einstein equations, see~\cite{curved} (see also~\cite{dw} for similar domain wall solutions in four-dimensional spacetimes).} 

In this paper, we point out some important phenomenological features of a modulus that is stabilized by a bulk scalar such as that of~\cite{GW2}.  If the large value $\krc\sim 12$ needed to solve the hierarchy puzzle arises from a small bulk scalar mass then the modulus potential that results is nearly flat (near its minimum) for values of the modulus VEV that solve the hierarchy problem in the manner of~\cite{RS1}.  As a consequence the modulus is likely to be lighter than the Kaluza-Klein modes of any  bulk field, and may be the first experimental signal for a scenario such as the Randall-Sundrum model.  In addition, its couplings to fields confined to the visible brane are suppressed by the TeV scale and are completely fixed by four-dimensional general covariance on the brane.  This leads to a well-defined set of predictions that can be compared with experiment.

First we review the modulus stabilization mechanism proposed in~\cite{GW2}.  To generate a potential for $\varphi$ we include a scalar field with bulk action
\begin{equation}
S_b={1\over 2}\int d^4 x\int_{-\pi}^\pi d\phi \sqrt{G} \left(G^{AB}\partial_A \Phi \partial_B \Phi - m^2 \Phi^2\right),
\end{equation}
where $G_{AB}$ with $A,B=\mu,\phi$ is given by Eq.~(\ref{eq:fluc}).  We also include interaction terms on the hidden and visible branes (at $\phi=0$ and $\phi=\pi$ respectively) given by
\begin{equation}
S_h = -\int d^4 x \sqrt{-g_h}\lambda_h \left(\Phi^2 - v_h^2\right)^2,
\end{equation}
and
\begin{equation}
S_v = -\int d^4 x \sqrt{-g_v}\lambda_v \left(\Phi^2 - v_v^2\right)^2,
\end{equation}
where $g_h$ and $g_v$ are the determinants of the induced metric on the hidden and visible branes respectively.  Note that $\Phi$ and $v_{v,h}$ have mass dimension $3/2$, while $\lambda_{v,h}$ have mass dimension $-2.$  As in~\cite{GW2}, we ignore the backreaction of $\Phi$ and $S_{v,h}$ on the spacetime geometry.  See~\cite{DFGK} for a treatment that includes these effects.

The terms on the branes cause $\Phi$ to develop a $\phi$-dependent vacuum expectation value $\Phi(\phi)$ which is determined classically by solving the equations of motion.  Inserting this solution into the bulk scalar action and integrating over $\phi$ yields an effective potential for $\varphi$ which has the form\begin{equation}
\label{eq:v}
V(\varphi)=\frac{k^3}{144 M^6} \varphi^4 \left(v_v - v_h (\varphi/f)^\epsilon\right)^2, 
\end{equation}
where by assumption $\epsilon\equiv m^2/4 k^2\ll 1$ and for simplicity terms of order $\epsilon$ have been neglected.  This potential has a minimum at 
\begin{equation}
\label{eq:min}
\frac{\langle \varphi\rangle}{f} = \left(\frac{v_v}{v_h}\right)^{1/\epsilon},
\end{equation}
or 
\begin{equation}
kr_c =k\langle T\rangle=\frac{1}{\pi\epsilon}\ln\left(v_h/v_v\right).
\end{equation}
In deriving Eq.~(\ref{eq:v}), we have taken the limit in which $\lambda_{v,h}$ are large.  This is done purely for convenience and the finite $\lambda$ case does not alter our conclusions significantly.  In particular, the leading $1/\lambda$ corrections to Eq.~(\ref{eq:v}) do not change the location of the minimum.  Note also that if $\ln (v_h/v_v)$ is of order unity, we only need $m^2/k^2$ of order $1/10$ to get $\krc\sim 12.$  Clearly, no extreme fine tuning of parameters is required to get the right magnitude for $\krc.$  For instance, taking  $v_h/v_v=1.5$ and the small bulk scalar mass $m/k = 0.2$ yields $\krc\simeq 12.$

From Eq.~(\ref{eq:v}), we can find the mass of $\varphi$ excitations about the minimum\footnote{Terms of order $\epsilon$ in the potential for $\varphi$ (which for simplicity have been neglected) give a contribution to $m_\varphi^2$ of order $\epsilon$ rather than order $\epsilon^2$ when treated as a perturbation.  Obviously they cannot really be neglected.  A correct treatment of these terms gives that $m_\varphi^2$ is suppressed from the TeV scale by $\epsilon^{3/2}$ rather than $\epsilon^2.$}:
\begin{equation}
\label{eq:mass}
m^2_\varphi =\frac{\partial^2 V}{\partial \varphi^2}(\langle\varphi\rangle)= \frac{k^2 v_v^2}{3 M^3}\epsilon^2 e^{-2\krcp}.
\end{equation}
Note that the exponential factor rescales $m_\varphi$ from a quantity of order the Planck scale down to the TeV scale.  Low-lying Kaluza-Klein excitations of bulk fields in the Randall-Sundrum model have masses which are typically slightly larger than the TeV scale~\cite{GW1,gauge}.  (This also includes the lowest excitation of the scalar $\Phi.$  Although it has a bulk mass which is smaller than the Planck mass, its lowest Kaluza-Klein mode still has a mass which is on the order of a few TeV~\cite{GW1}.)  However if the large value of $kr_c$ (i.e, $kr_c\sim 12$) arises from a small bulk scalar mass then in addition to the  factor $\exp(-2 \krc)$ in Eq.~(\ref{eq:mass}) there is suppression by the small quantity $\epsilon$.  Consequently, $m_\varphi$ is somewhat smaller than the TeV scale, and therefore lighter than the Kaluza-Klein excitations of bulk fields\footnote{In this case, the backreaction of the scalar field on the five-dimensional metric is small (i.e. suppressed by powers of $m/k$) and the mixing of the modulus with Kaluza-Klein excitations of the graviton and $\Phi$ are expected to be small.}.  Detection of the radion $\varphi$ might be the first clear signal of the scenario of~\cite{RS1}.

Because the radion arises as a gravitational degree of freedom, its couplings to brane matter are constrained by four-dimensional general covariance.  These couplings arise from the induced metric on the brane.  On the $\phi=0$ brane, the induced metric obtained from Eq.~(\ref{eq:fluc}) is simply $g_{\mu\nu}$:  the modulus does not couple directly to hidden brane matter.  The induced metric on the visible brane is given by $(\varphi/f)^2 g_{\mu\nu}$ and consequently, $\varphi$ interacts directly with Standard Model fields.  For example, consider a scalar $h(x)$ confined to the visible brane:
\begin{equation}
S=\frac{1}{2}\int d^4 x \sqrt{-g} (\varphi/f)^4 [(\varphi/f)^{-2} g^{\mu\nu} \partial_\mu h \partial_\nu h - \mu_0^2 h^2].
\end{equation}
Rescaling $h\rightarrow (f/\langle\varphi\rangle) h$ to obtain a canonically normalized field, this becomes
\begin{equation}
\label{eq:couple}
S=\frac{1}{2}\int d^4 x \sqrt{-g}[(\varphi/\langle\varphi\rangle)^2 g^{\mu\nu} \partial_\mu h \partial_\nu h - \mu^2(\varphi/\langle\varphi\rangle)^4 h^2],
\end{equation}
where 
\begin{equation}
\mu=\mu_0 \frac{\langle\varphi\rangle}{f} = \mu_0 e^{-\krcp}.
\end{equation}
For $\mu_0$ of order the Planck scale and $\krc\sim 12,$ the physical mass $\mu$ is of order the weak scale.  This result can be generalized to any operator appearing in the visible brane Lagrangian:  a parameter with mass dimension $d$ is rescaled by $d$ powers of $\exp{(-\krcp)}$.  Also, any operator with $n$ powers of the inverse metric is multiplied by $4-2n$ powers of $\varphi/\langle\varphi\rangle$ (for fermions, a power of the inverse vierbein counts as $n=1/2$).  Note that the couplings of the modulus $\varphi$ to visible brane fields are characterized by the scale $\langle\varphi\rangle,$ which is in the TeV range.  Expanding $\varphi$ about its VEV, $\varphi=\langle\varphi\rangle + \delta\varphi$, we see that $\delta\varphi$ couples to ordinary matter through the trace of the Standard Model energy-momentum tensor $T_{\mu\nu},$
\begin{equation}
\label{eq:trace}
{\cal L}_{int} = \frac{\delta\varphi}{\langle\varphi\rangle} T^\mu{}_\mu.
\end{equation}
Neglecting the quark masses, the energy momentum tensor for QCD is traceless at tree level.  This suppresses some production mechanisms for the radion at high energy hadron colliders.

The couplings of the radion to Standard Model fields are like those of the Higgs particle.  In the Standard Model, the latter has a coupling to the Z boson given by
\begin{equation}
{\cal L}_{int} = \frac{e}{\sin 2\theta_W} m_Z h Z^\mu Z_\mu,
\end{equation}
where $h$ is the canonically normalized neutral Higgs scalar.  On the other hand the analogous radion coupling is given by
\begin{equation}
{\cal L}_{int} = \frac{\delta\varphi}{\langle\varphi\rangle} m_Z^2 Z^\mu Z_\mu.
\end{equation}
Consequently, radion production by a virtual Z is suppressed relative to the analogous Higgs production process by $(m_Z\sin 2\theta_W/e\langle\varphi\rangle)^2.$

If the almost complete cancellation of the $\phi$ integral of the first two terms in the square brackets of Eq.~(\ref{eq:phi}) did not occur, the canonically normalized modulus field would have couplings suppressed by the Planck scale instead of the weak scale, as well as a much lighter mass, of order $\,(\mbox{TeV})^2/M_{Pl}$.  In this case, the phenomenology of the field $\varphi$ would be similar to that of the radion that arises in scenarios with large extra dimensions~\cite{adm}.  It would be interesting to use the methods of~\cite{DFGK} to examine precisely how deviations from the pure anti-deSitter metric of Eq.~(\ref{eq:metric}), which arise due to the classical $\Phi$ configuration, influence the kinetic term for $T.$

We have explored some of the physical properties of the radion field which arises in the Randall-Sundrum scenario.  An important feature is that it couples to visible brane matter with TeV rather than Planck scale strength.  In the absence of a mechanism that generates a modulus mass, this is unacceptable:  modulus exchange gives rise to a long range universal attractive force which is 32 orders of magnitude stronger than gravity.  On the other hand, this is not a problem if the radion is stabilized by a mechanism such as that of~\cite{GW2}.  In addition, if the large value of $\krc\sim 12$ arises from a small bulk scalar mass, then the stabilized modulus has a very distinctive phenomenology.  It has a mass which is lighter than Kaluza-Klein modes of bulk fields, making it the first direct signal of the extra dimension.  Also, the couplings of the modulus to the Standard Model fields are fixed by general covariance and depend on the single parameter $\langle\varphi\rangle.$  We expect similar phenomenology to arise in other scenarios, such as those of~\cite{others,LR}.  Given this definite pattern of $\varphi$ couplings, it would be worthwhile to consider constraints from high energy accelerator experiments.  It may also be interesting to explore the role of this field in a cosmological setting.

Finally, we note that in~\cite{DFGK} other regions of parameter space which generate a large value of $\krc$ were explored.  For example, $\krc\sim 12$ can be obtained if the bulk scalar has negative mass squared and its VEV on the visible brane is large compared with that on the hidden brane.  It is possible that even in these cases, $\varphi$ will be light for large $\krc$ since a natural way to get a large VEV for $T$ (i.e., a large value of $\krc$) is to have its potential be broad.  In these regions of parameter space, the backreaction of the vacuum $\Phi$ configuration on the five-dimensional metric is not small.  When the backreaction is included, the induced metric will have a more complicated dependence on $\varphi$ than $(\varphi/f)^2 g_{\mu\nu}$.  There might also be significant mixing with the Kaluza-Klein modes of the metric and of $\Phi,$ in which case the coupling of the physical mass eigenstates to matter will be more complicated than the situation we have considered.  These points need to be examined before one can have confidence that $\varphi$ is generically lighter than the TeV scale.   Also, there is the issue of whether the lightness of $\varphi$ will survive quantum corrections.  However, it seems unlikely that this question can be definitively addressed with field theory methods alone. 

This work was supported in part by the Department of Energy under grant number DE-FG03-92-ER 40701.

{\bf Note added}  While this work was in preparation, \cite{RB} appeared which contains some results that are similar to those presented here.


\begin{references}

\bibitem{SUSY} E. Witten, Nucl. Phys. B188 (1981) 513.

\bibitem{tech} L. Susskind, Phys. Rev. D20 (1979) 2619; S. Weinberg, Phys. Rev. D13 (1976) 974; Phys. Rev. D19 (1979) 1277.

\bibitem{xdim}  N. Arkani-Hamed, S. Dimopoulos, and G. Dvali, Phys. Lett. B429 (1998) 263; I. Antoniadis, N. Arkani-Hamed, S. Dimopoulos, and G. Dvali, Phys. Lett. B436 (1998) 257.

\bibitem{RS1} L. Randall and R. Sundrum, Phys. Rev. Lett. 83 (1999) 3370.

\bibitem{HW} P. Horava and E. Witten, Nucl. Phys. B460 (1996) 506; E. Witten, Nucl. Phys. B471 (1996) 135; P. Horava and E. Witten, Nucl. Phys. B475 (1996) 94.

\bibitem{mtheory} A. Lukas, B.A. Ovrut, and D. Waldram, hep-th/9806022; hep-th/9902071.

\bibitem{supergrav} A. Kehagias, hep-th/9906204; J. Cline, C. Grojean, and G. Servant, hep-th/9910081.

\bibitem{verlinde} H. Verlinde, hep-th/9906182.

\bibitem{others} I. Oda, hep-th/9908104; T. Li, hep-th/9908174; K.R. Dienes, E. Dudas, and T. Ghergetta, hep-ph/9908530; H. Hatanaka, M. Sakamoto, M. Tachibana, and K. Takenaga, hep-th/9909076.

\bibitem{LR} J. Lykken and L. Randall, hep-th/9908076.

\bibitem{RS2} L. Randall and R. Sundrum, hep-th/9906064.

\bibitem{DHR1} H. Davoudiasl, J.L. Hewett, and T.G. Rizzo, hep-ph/9909255. 

\bibitem{GW1} W.D. Goldberger and M.B. Wise, Phys. Rev. D60 (1999) 107505.

\bibitem{gauge} H. Davoudiasl, J.L. Hewett, and T.G. Rizzo, hep-ph/9911262; A. Pomarol, hep-ph/9911294.  

\bibitem{GW2} W.D. Goldberger and M.B. Wise, Phys. Rev. Lett. 83 (1999) 4922.

\bibitem{DFGK} O. De Wolfe, D.Z. Freedman, S.S. Gubser and A. Karch, hep-th/9909134.

\bibitem{curved} T. Nihei, hep-ph/9905487;  N. Kaloper, hep-th/9905210.

\bibitem{dw} M. Cvetic, S. Griffies, and H.H. Soleng, Phys. Rev. D48 (1993) 2613; M. Cvetic, S. Griffies, and S. Rey, Nucl. Phys. B381 (1992) 301; M. Cvetic and H. Soleng, Phys. Rept. 282:159-223 (1997). 

\bibitem{adm} N. Arkani-Hamed, S. Dimopoulos, and J. March-Russell, hep-th/9809124.

\bibitem{RB}  C. Csaki, M. Graesser, L. Randall, and J. Terning, hep-ph/9911406.

\end{references}
\end{document}